\documentclass{webofc}
\usepackage[varg]{txfonts}   % Web of Conferences font
\usepackage{xpunctuate}
\usepackage{subfig}
\usepackage{siunitx}
\usepackage[noabbrev]{cleveref}
\usepackage{csquotes}

\newcommand*{\ie}{i.e\xperiod}

\newcommand*{\qqbarPrime}{\ensuremath{q\bar{q}'}\xspace}
\newcommand*{\JP}{\ensuremath{J^{P}}\xspace}

\DeclareSIUnit{\clight}{\text{\ensuremath{c}}}
\DeclareSIUnit[per-mode=symbol]\MeVcc{\MeV\per\clight\squared}
\DeclareSIUnit[per-mode=symbol]\GeVcc{\GeV\per\clight\squared}
\DeclareSIUnit[per-mode=symbol]\GeVc{\GeV\per\clight}
\DeclareSIUnit[per-mode=symbol]\GeVcsq{(GeV/\clight)^2}  

 \providecommand*{\abs}[1]{}                                                                                                                                 
 \renewcommand*{\abs}[1]{\ensuremath{\vert{#1}\vert}\xspace}
  
 \newcommand*{\Kpipi}{\ensuremath{{K^-\pi^-\pi^+}}\xspace}      
 \newcommand*{\Kpi}{\ensuremath{{K^-\pi^+}}\xspace}   
 \newcommand*{\pipi}{\ensuremath{{\pi^-\pi^+}}\xspace}   
 \newcommand*{\mTwoPi}{\ensuremath{m_\pipi}\xspace}

 \newcommand*{\mKpipi}{\ensuremath{m_{K\pi\pi}}\xspace}  
 \newcommand*{\tpr}{\ensuremath{{t'}}\xspace}

 \newcommand*{\WaveK}[7]{\ensuremath{{#1}^{{#2}}\,\allowbreak{#3}^{#4}\,\allowbreak{#5}\,{#6}\,{#7}}\xspace}

\newcommand{\threeFigureSubfigureWidth}{0.333\linewidth}
\newcommand{\twoFigureSubfigureWidth}{0.5\linewidth}  
\begin{document}
\title{Strange-Meson Spectroscopy -- from COMPASS to AMBER}

\author{\firstname{S.} \lastname{Wallner}\inst{1}\fnsep\thanks{\email{swallner@mpp.mpg.de}}
\firstname{for the COMPASS and AMBER collaborations}
        % etc.
}

\institute{Max-Planck-Institut für Physik, 80805 Munich, Germany
          }

\abstract{%
  COMPASS is a multi-purpose fixed-target experiment at CERN's M2 beam line aimed at studying the structure and spectrum of hadrons.
  It has collected the so far world's largest data set on diffractive production of the \Kpipi final state, which in principle gives access to all strange mesons.
  Based on this data set, we performed an elaborate partial-wave analysis.
  It reveals signals in the mass region of well-known states, such as the $K_2^*(1430)$.
  In addition, we found indications for a resonance-like signal in the mass region of the $K(1630)$. This state would be a supernumerary state and hence could be a candidate for an exotic strange meson. The partial-wave analysis is limited in some areas by the limited kinematic coverage of the final-state particle identification of the COMPASS setup. To overcome this limitation, we propose a new high-precision strange-meson spectroscopy measurement at the AMBER experiment, which will be located at CERN's M2 beam line.
}
\maketitle
\section{Introduction}

\begin{figure}[h]
	\centering%
	\includegraphics[width=\linewidth]{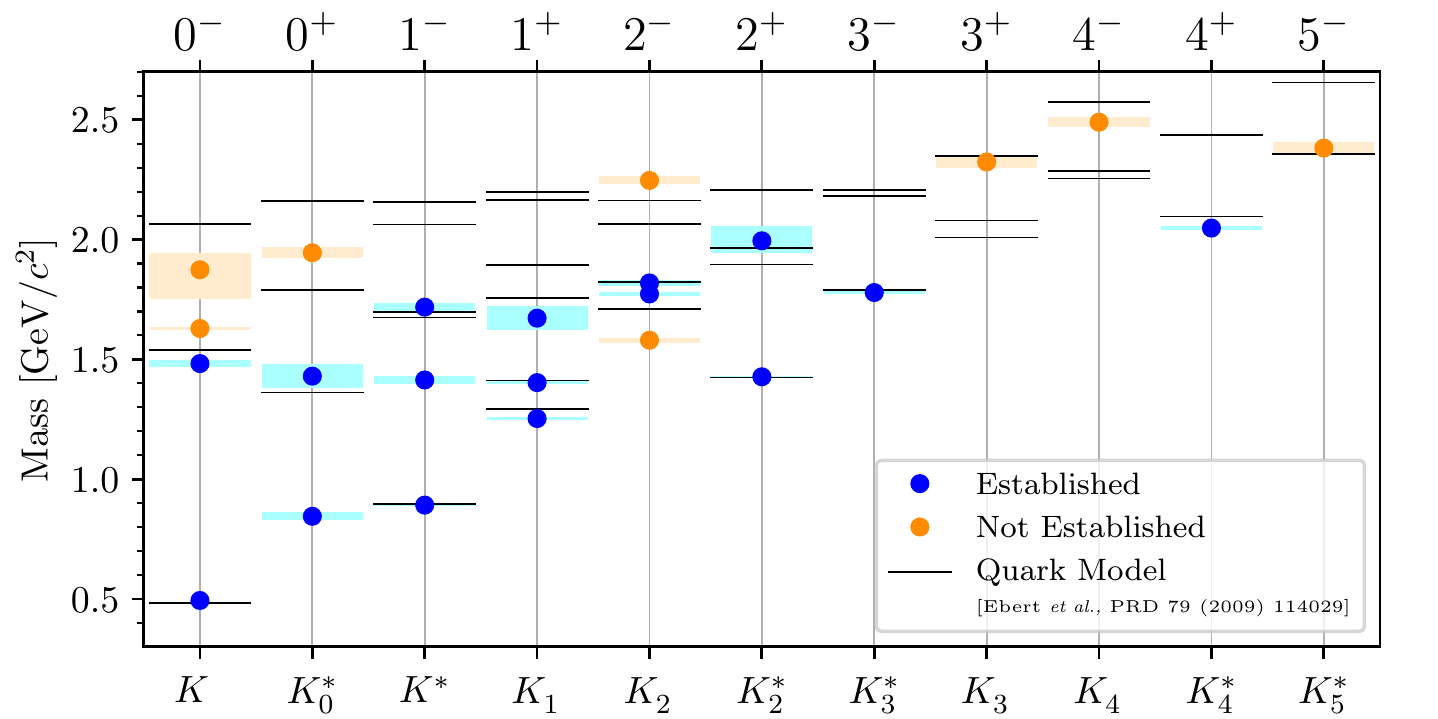}%
	\caption{Strange-meson spectrum grouped by the $J^{P}$	quantum numbers of the states. The blue data points show the masses of established states, the orange data points those of not established states as listed by the PDG~\cite{Zyla2021}. The similarly colored boxes represent the corresponding uncertainties. The black horizontal lines show the masses of states as predicted by the quark-model calculation in ref.~\cite{Ebert2009}.}
	\label{fig:kaonspectrum}
\end{figure}

In order to establish a complete picture of the light-meson sector, an important goal is to find all the strange partners of the non-strange light mesons, \ie to complete the corresponding SU(3) flavor nonets. This includes the search for exotic mesons beyond pure quark-model \qqbarPrime states.
Furthermore, strange mesons play an important role in other fields of particle physics. For example, in searches for $CP$ violation, where $B$- or $D$-meson decays to multi-body hadronic final states containing kaons are studied. In these decays, excited strange mesons appear as intermediate resonances.
Hence, a complete knowledge of the spectrum of strange mesons is important to precisely understand these decays.
However, compared to the non-strange light-meson spectrum, many parts of the strange-meson spectrum are still poorly explored. Figure~\ref{fig:kaonspectrum} shows our current knowledge of the strange-meson spectrum according to the PDG~\cite{Zyla2021}. In total, only 25 strange mesons are listed and only 16 of them are considered as established states (blue points). Especially in the high-mass region, many strange-meson candidates (orange) need further confirmation and many states predicted by quark-models (black lines) are missing. This is because at higher masses the states typically become broader, which leads to large overlaps among them. Hence, they are experimentally more challenging to extract.
%\clearpage

%=============================================================================
%=============================================================================
%=============================================================================
\section{Strange-Meson Spectroscopy at COMPASS}
COMPASS is a fixed-target multi-purpose experiment located at CERN.
Positive or negative secondary hadron beams or a tertiary muon beam are directed onto various types of targets.
Two CEDAR detectors before the target allow to identify the beam particle species.
The forward-going final-state particles are measured by a two-stage magnetic spectrometer and are identified by a ring-imaging Cherenkov detector~\cite{Abbon:2014aex}.
So far, COMPASS has studied mainly isovector resonances of the $a_J$ and $\pi_J$ families with high precision, using the dominant $\pi^-$ component of the negative hadron beam, which has a momentum of  \SI{190}{\GeVc}~\cite{Adolph2015,Akhunzyanov2018}.
In this analysis, we study the strange-meson spectrum up to masses of \SI{3}{\GeVcc} using the \SI{2.4}{\percent} $K^-$ component of the beam.
%Due to their very short lifetime, we observe the resonances only in their decays into quasi-stable final-state particles.
Our flagship channel is the $K^-\pi^-\pi^+$ final state, which is produced in diffractive scattering off a liquid-hydrogen target and in principle gives access to all strange-meson states, i.e. $K_J$ and $K^*_J$ mesons. %\footnote{Except for $J^P = 0^+$ states, which are not produced in diffractive scattering.}
COMPASS acquired the so far world's largest data set of about \num{720000} exclusive events for this final state.

\begin{figure}%
	\centering%
	\subfloat[]{\includegraphics[width=\threeFigureSubfigureWidth]{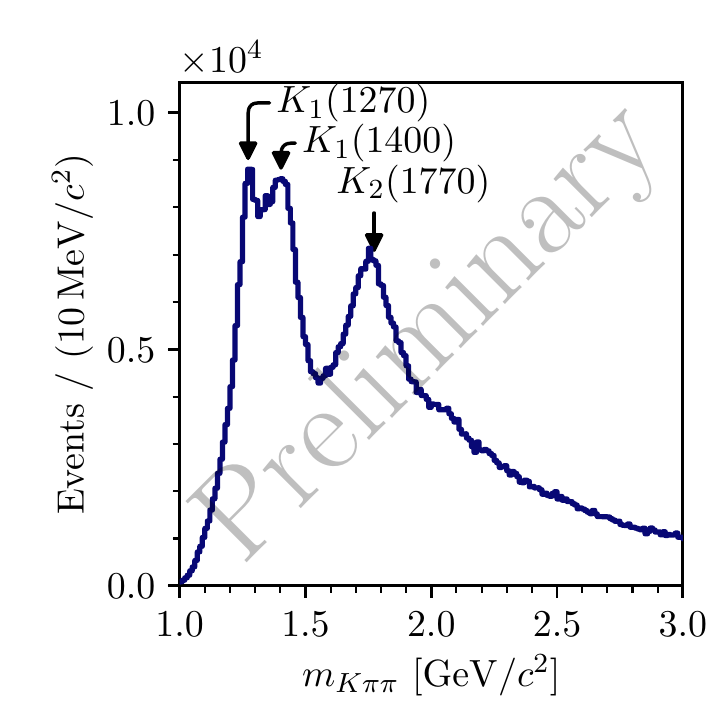}\label{fig:kin:mKpipi}}%
	\subfloat[]{\includegraphics[width=\threeFigureSubfigureWidth]{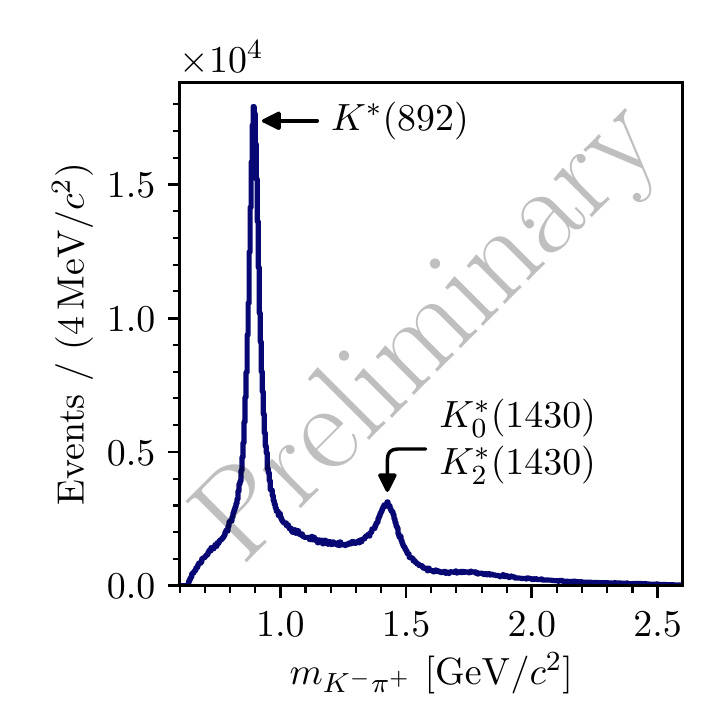}\label{fig:kin:mKpi}}%
	\subfloat[]{\includegraphics[width=\threeFigureSubfigureWidth]{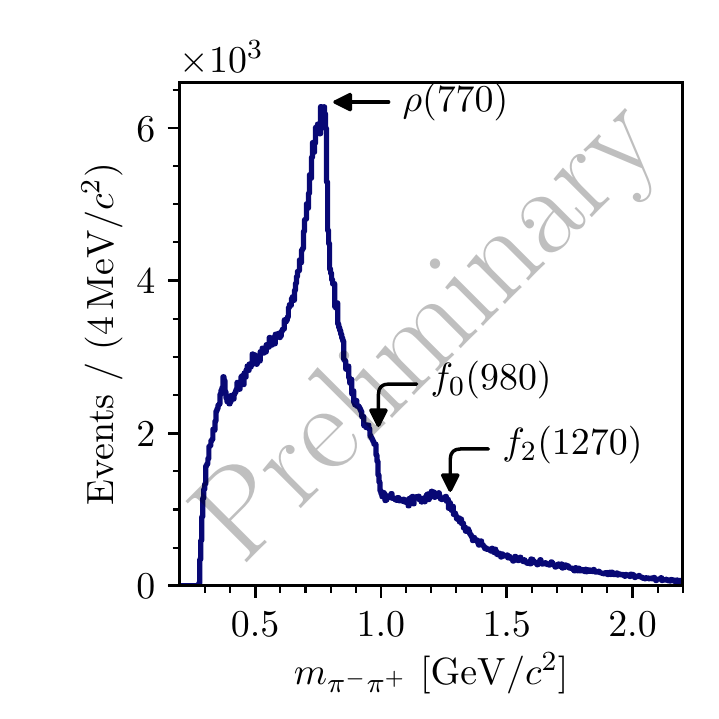}\label{fig:kin:mpipi}}%
	\caption{Invariant mass distribution of (a) the \Kpipi system, (b) the \Kpi subsystem, and (c) the \pipi subsystems. The distributions are not corrected for acceptance effects, which distort these distributions.}%
	\label{fig:kin}%
\end{figure}

\Cref{fig:kin:mKpipi} shows the invariant mass spectrum of the \Kpipi system. It exhibits structures in the \mKpipi region of well-known resonances, e.g. $K_1(1270)$ and $K_1(1400)$, which sit on top of a broad distribution.
The invariant mass distributions of the \Kpi and \pipi subsystems are dominated by well-known two-body resonances as indicated by the labels in \cref{fig:kin:mKpi,fig:kin:mpipi}.\footnote{The small spike at about \SI{0.4}{\GeVcc} in the \mTwoPi system originates from $\phi \to K^-K^+$ decays from a small contamination of the \Kpipi sample by $K^-K^-K^+$ events, where the kaons are misidentified as pions.}
This dominance of two-body resonances justifies the application of the isobar model in the partial-wave decomposition, which is discussed in the following section.

%=============================================================================
%=============================================================================
%=============================================================================
\section{Analysis Method: Partial-Wave Decomposition}
We employ the method of partial-wave analysis to decompose the data into contributions from various partial waves with well-defined quantum numbers (see ref.~\cite{Adolph2015} for details).
To this end, we construct a model for the intensity distribution $\mathcal{I}(\tau)$ of the events as a function of the five \Kpipi phase-space variables that are represented by $\tau$. $\mathcal{I}(\tau)$ is modeled as a coherent sum of partial-wave amplitudes, \ie
\begin{equation}
\mathcal{I}(\tau; \mKpipi, \tpr) = \left|\sum\limits_{a}^{\text{waves}} {\mathcal T_a(\mKpipi, \tpr)}\, {\psi_a(\tau;\mKpipi)}\right|^2. \label{eq:intensity}
\end{equation}
In the isobar model, these partial waves are represented by $a = \WaveK{J}{P}{M}{\varepsilon}{\zeta}{b}{L}$, which is defined by the $J^PM^\varepsilon$ quantum numbers of the \Kpipi system,\footnote{Here, $J$ is the spin of the \Kpipi state and $P$ its parity. The spin projection along the beam axis is expressed in the reflectivity basis~\cite{Chung1975} and given by $M^\varepsilon$, where $\varepsilon$ is the reflectivity given by the naturality of the exchange particle.} the intermediate two-body resonance $\zeta$,\footnote{We consider isobar resonances in the \Kpi and \pipi subsystem.} and the orbital angular momentum $L$ between the bachelor particle $b$ and the isobar.
Within the isobar model, the decay amplitudes $\psi_a$ can be calculated. This allows us to extract the partial-wave amplitudes $\mathcal T_a$, which determine strength and phase of each wave from the data by an unbinned maximum-likelihood fit.

In order to extract the dependence of the partial-wave amplitudes on the invariant mass \mKpipi of the \Kpipi system and on the reduced squared four-momentum transfer \tpr between the beam particle and the target proton, the partial-wave decomposition fit is performed independently in 75 narrow bins of \mKpipi and 4 bins of \tpr in the analyzed kinematic range of $1.0 \leq \mKpipi < \SI{3.0}{\GeVcc}$ and $0.1 \leq \tpr < \SI{1.0}{\GeVcsq}$.
%\footnote{By definition, \tpr is a positive quantity, as $\tpr\equiv \abs{t} - \abs{t}\sub{min}$.}
%In addition, the size of the dataset allows us to study the dependence of the partial-wave amplitudes on the squared four-momentum transfer \tpr, by binning the data in \tpr as well.
%By performing the partial-wave decomposition independently in $75$ \mKpipi bins in the range $1.0 < \mKpipi < \SI{3.0}{\GeVcc}$ and $4$ \tpr bins in the range $0.1 < \tpr < \SI{1.0}{\GeVcsq}$,\footnote{By definition, \tpr is a positive quantity, as $\tpr\equiv \abs{t} - \abs{t}\sub{min}$.} we extract simultaneously the \mKpipi and \tpr dependence of the partial-wave amplitudes from the data.

In order to construct the wave set, \ie the set of partial waves that enter the sum in \cref{eq:intensity}, we apply model-selection techniques.
To this end, we systematically construct a large set of possible partial waves, called wave pool.
We allow for all combinations within the following loose restrictions: $J\leq 7$; $L\leq 7$; positive naturality of the exchange particle\footnote{We consider only waves with $\varepsilon=+$, as we assume positive naturality of the exchange particle (see ref.~\cite{Ketzer2020a}).}; and twelve isobars: two $K\pi$ $S$-wave amplitudes, $K^*(892)$, $K^*(1680)$, $K^*_2(1430)$, $K^*_3(1780)$, a broad $\pi\pi$ $S$-wave amplitude, $f_0(980)$, $f_0(1500)$, $\rho(770)$, $f_2(1270)$, and $\rho_3(1690)$.\footnote{We use relativistic Breit-Wigner amplitudes for all isobar line shapes, except for the $S$-wave isobars, where we employ $K$-matrix parameterizations (see ref.~\cite{Wallner:2022prx}).}
This results in a wave pool of 596 allowed waves.
In order to select the wave set, we fit the wave pool to the data applying regularization techniques to suppress insignificant waves (see ref.~\cite{Kaspar:2019qpf} for details).
This results in an individual wave set for each $(\mKpipi, \tpr)$ cell, which we fit again to the data without applying regularization terms in order to obtain our final results, which are discussed in the following sections.

%=============================================================================
%=============================================================================
%=============================================================================
\section{Partial Waves with $\JP=2^+$}

\begin{figure}%
	\centering%
	\subfloat[]{\includegraphics[width=\threeFigureSubfigureWidth]{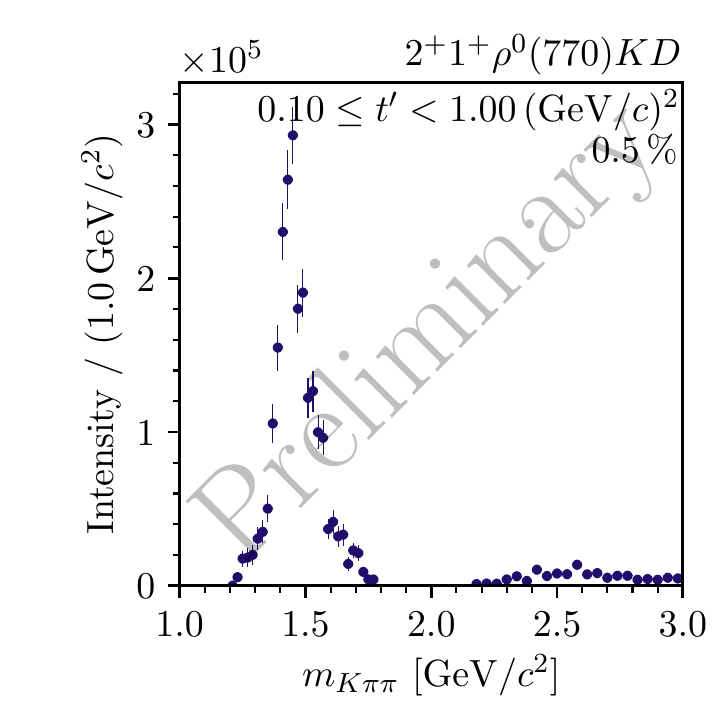}\label{fig:2p:rho}}%
	\subfloat[]{\includegraphics[width=\threeFigureSubfigureWidth]{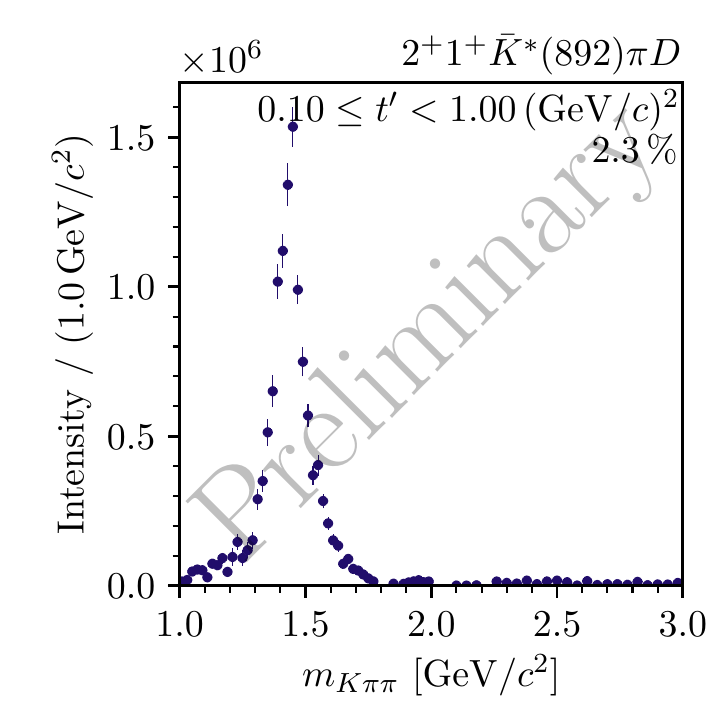}\label{fig:2p:KMst}}%
	\subfloat[]{{\includegraphics[width=\threeFigureSubfigureWidth]{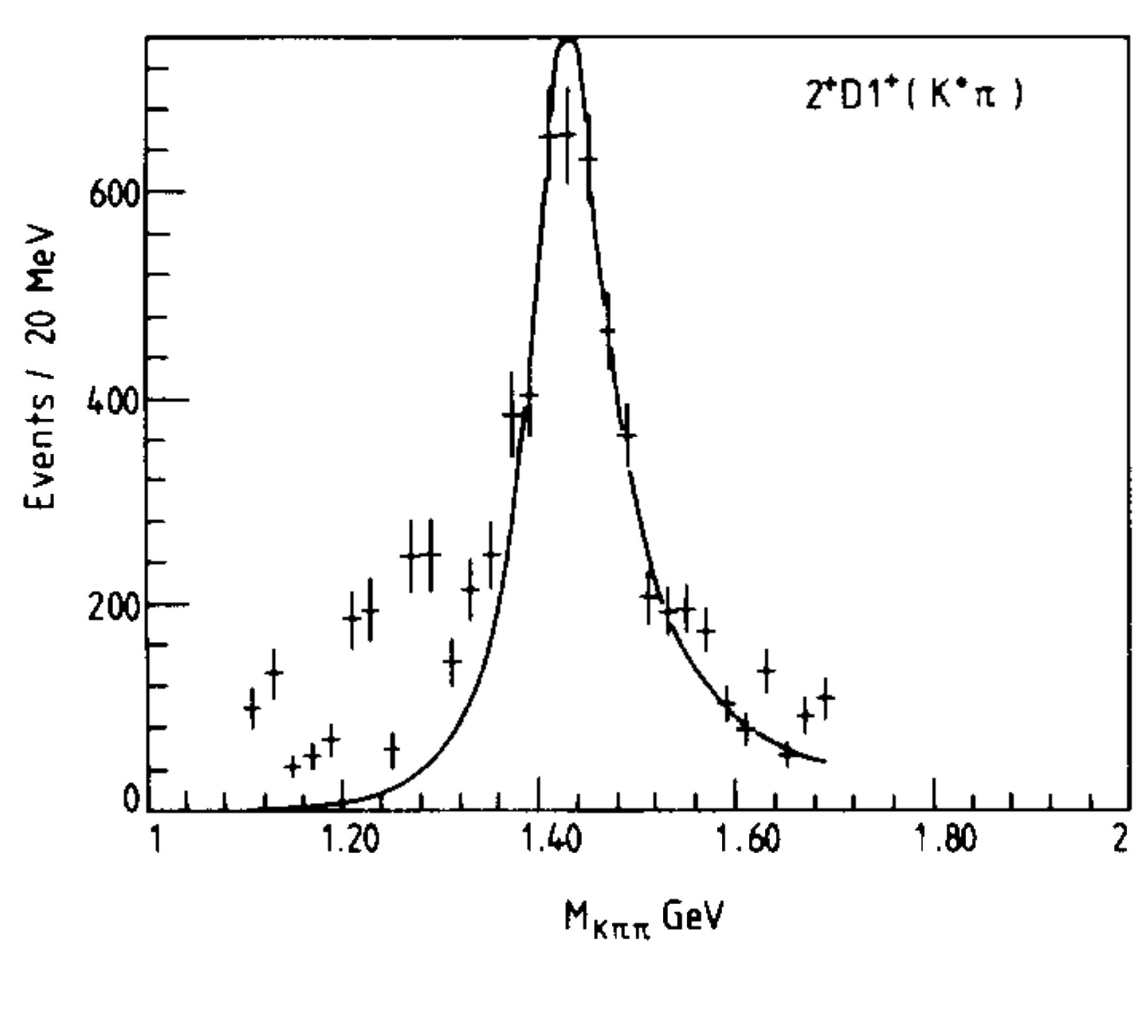}\label{fig:2p:ACCMOR}}}%
	\caption{Intensity distributions of partial waves with $\JP=2^+$. (a) and (b) show the intensity distributions of the \WaveK2+1+{\rho(770)}KD and \WaveK2+1+{K^{*}(892)}\pi D waves from our analysis in the range $0.1\leq \tpr < 1.0\,\si{\GeVcsq}$. The percentage numbers in the upper-right corners of (a) and (b) give the relative intensity of the corresponding wave. (c) shows the intensity distribution of the \WaveK2+1+{K^{*}(892)}\pi D  wave as obtained in the analysis of WA03 data performed by the ACCMOR collaboration~\cite{Daum1981a} in the range $0.05 \leq \tpr \leq 0.7\,\si{\GeVcsq}$. The curve represents a resonance model that was fitted to the data.}%
	\label{fig:2p}%
\end{figure}

The partial waves with $\JP=2^+$ exhibit one of the clearest signals in this channel.
\Cref{fig:2p:rho} shows the intensity of the \WaveK2+1+{\rho(770)}KD wave as a function of \mKpipi. This partial wave represents the production and decay of a state with $\JP=2^+$, decaying into $\rho(770)$ and $K$ in a $D$ wave.
The intensity distribution of this wave exhibits a clear peak at about \SI{1.4}{\GeVcc}. This peak can be attributed to the well known $K_2^*(1430)$ resonances.

The \Kpipi final state allows us to study various decays modes in a single data set. For example, $\JP=2^+$ states can also decay into $K^*(892)$ and $\pi$. This decay mode is represented by the \WaveK2+1+{K^{*}(892)}\pi D wave. \Cref{fig:2p:KMst} shows the corresponding intensity distribution, which also exhibits a clear $K_2^*(1430)$ peak at about \SI{1.4}{\GeVcc}.

For comparison, \cref{fig:2p:ACCMOR} shows the intensity distribution of the same partial wave as obtained in an analysis performed by the ACCMOR collaboration~\cite{Daum1981a}. This analysis is based on the perviously world's largest sample on the diffractively produced \Kpipi final state of about \num{200000} events. The signal observed by ACCMOR in this wave is consistent with our findings. However, the COMPASS data appears to have less background, in particular in the low-mass region.

The clear observation of a resonance-like signal consistent with the well-know $K_2^*(1430)$ and the high purity of this signal validates our analysis results and demonstrates the good quality of our data.

%=============================================================================
%=============================================================================
%=============================================================================
\section{Exotic Strange Mesons}
\label{sec:exotic}

Today, one of the major goals of meson spectroscopy is to establish the existence of exotic states, \ie states beyond \qqbarPrime quark-model sates. In the strange-meson sector, such states are expected to show up as supernumerary states in addition to the predicted quark-model states. Hence, identifying exotic strange mesons requires mapping out the complete strange-meson spectrum over a wide mass range in a consistent way.

An interesting sector to search for exotic strange mesons is the $\JP=0^-$ sector. 
In the mass region below \SI{2.5}{\GeVcc}, the PDG~\cite{Zyla2021} lists at the moment the $K(1460)$, which is considered as an established state, and the $K(1830)$, which needs further confirmation. In addition, the PDG lists the $K(1630)$. Although the PDG lists this state as $K(1630)$, its quantum numbers are actually undetermined. Also, so far the $K(1630)$ signal was observed by only a single experiment with an unexpectedly small width of only \SI{16}{\MeVcc}. This is much narrower than one would expect for a strange-meson resonance.
Quark-model calculations~\cite{Ebert2009} predict only two excited strange-mesons with $\JP=0^-$ in the mass region up to \SI{2.5}{\GeVcc}. The lighter predicted state agrees best with the $K(1460)$, while the heavier predicted state agrees best with the $K(1830)$ (see horizontal black lines in \cref{fig:kaonspectrum}).

\begin{figure}[]%
	\centering%
	\subfloat[]{\includegraphics[width=\twoFigureSubfigureWidth]{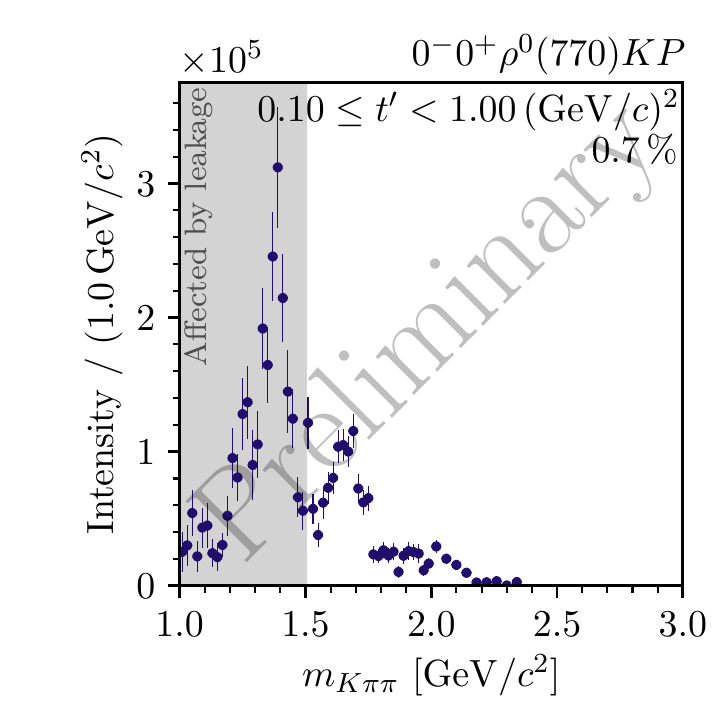}\label{fig:0m:int}}%
	\subfloat[]{\includegraphics[width=\twoFigureSubfigureWidth]{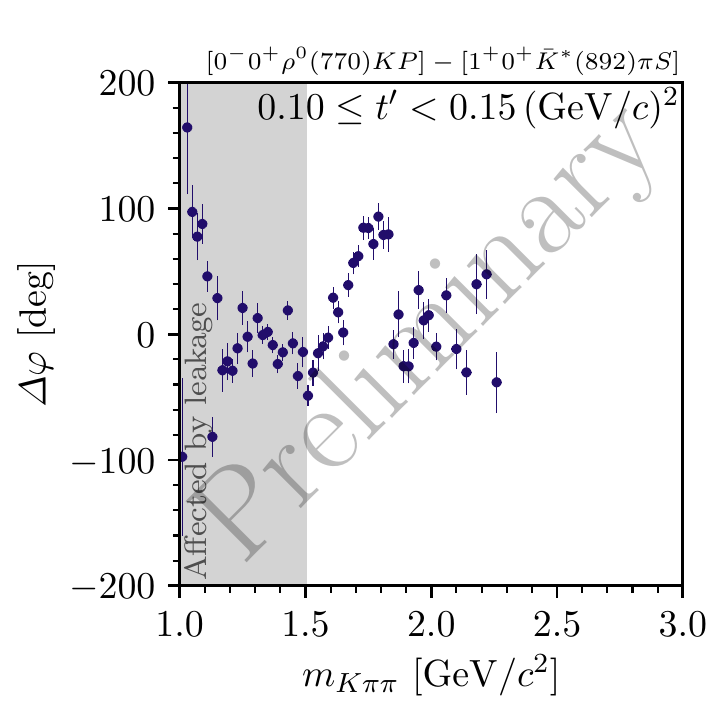}\label{fig:0m:phase}}%
	\caption{Intensity and relative phase of the \WaveK0-0+{\rho(770)}KP wave. (a) shows the intensity distribution in the range $0.1\leq \tpr < 1.0\,\si{\GeVcsq}$. (b) shows phase of the \WaveK0-0+{\rho(770)}KP wave with respect to the \WaveK1+0+{K^{*}(892)}\pi S  wave in the lowest analyzed \tpr bin, \ie in the range $0.10 \leq \tpr < 0.15\,\si{\GeVcsq}$. The percentage number in the upper-right corner of the intensity spectrum is the relative intensity
		of this wave.}%
	\label{fig:0m}%
\end{figure}

At COMPASS, we study strange mesons with $\JP=0^-$ in their decay via the $\rho(770)\,K\,P$ wave to the \Kpipi final state.
\Cref{fig:0m:int} shows the intensity distribution of the corresponding partial wave.
It exhibits a peak at about \SI{1.4}{\GeVcc}, which potentially is caused by the $K(1460)$ resonance. However, in the range $\mKpipi \lesssim \SI{1.5}{\GeVcc}$, this partial wave is affected by known analysis artifacts (see \cref{sec:leakage}). Hence, we cannot make robust statements about the $K(1460)$ from COMPASS data.
At about \SI{1.7}{\GeVcc}, the intensity distribution exhibits another clear peak. This peak is accompanied by a clear rise of the relative phase of this partial wave by about \SI{120}{\degree} as shown in \cref{fig:0m:phase}.
The peak in the intensity, as well as the rising phase, indicate a resonance-like signal in the \SI{1.7}{\GeVcc} mass region. This signal is too high in mass to be compatible with previous observations of the $K(1460)$ and to low in mass to be compatible with previous observations of the $K(1830)$. While the mass of our \SI{1.7}{\GeVcc} signal would be consistent with the mass observed for the $K(1630)$, we observe a width that is clearly larger than \SI{16}{\MeVcc}.
At about \SI{2}{\GeVcc}, the intensity distribution exhibits another weak potential signal that may arise from the $K(1830)$ resonance.

In total, we observe indications for three resonance-like signals in the \WaveK0-0+{\rho(770)}KP wave. The lightest and the heaviest of these three signals are in the mass ranges of the previously observed $K(1460)$ and $K(1830)$ and agree best with the two predicted quark-model states. 
This would mean that the $K(1630)$ signal, which we observe at about \SI{1.7}{\GeVcc} and which is actually the clearest signal in the \WaveK0-0+{\rho(770)}KP wave, is a supernumerary state making it a candidate for an exotic strange-meson resonance. 
However, further studies need to be carried out to investigate these signals. For example, the resonance contributions to this partial wave need to be modeled and fitted to the measured partial-wave amplitude in order to validate the resonance-like characteristic of the observed signals and to measure the corresponding mass and width parameters.

%=============================================================================
%=============================================================================
%=============================================================================
\section{Limitations for Strange-Meson Spectroscopy at COMPASS}
\label{sec:leakage}

At COMPASS, the momenta of final-state particles cover a large range reaching values up to the nominal beam momentum of about \SI{190}{\GeVc}. However, our final-state particle identification can separate kaons from pions only up to particle momenta of about \SI{50}{\GeVc}.
As analyzing the \Kpipi data requires to identify which of the two negative final-state particle is the pion and which is the kaon, we cannot use events where we cannot identify these particles. 
This leads to a blind spot in our experimental acceptance.

\begin{figure}%
	\centering%
	\includegraphics[width=\twoFigureSubfigureWidth]{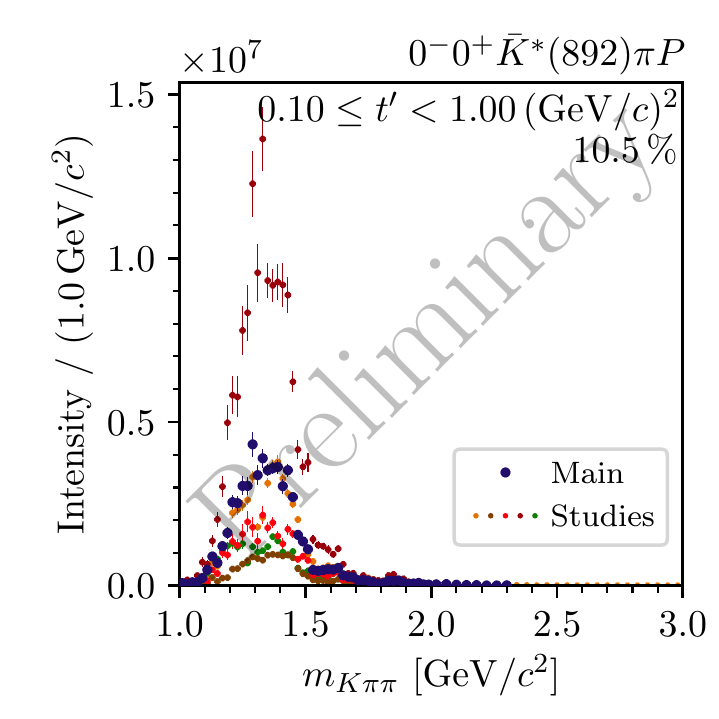}%
	\caption{Intensity distribution of the \WaveK0-0+{K^{*}(892)}\pi P  wave in the range $0.1\leq \tpr < 1.0\,\si{\GeVcsq}$. The large blue data points represent the result of the \enquote{main} analysis. The small data points show the results from five systematic studies, where different aspects of the analysis where changed. Each systematic study is shown by one color. The percentage number in the upper-right corner of the intensity spectrum is the relative intensity
	of this wave as obtained in the \enquote{main} analysis.}%
	\label{fig:leakage}%
\end{figure}

In the partial-wave decomposition, this blind spot causes a loss of distinguishing power among certain partial waves. This leads, for example, to artificially large intensities of these partial waves, which are sensitive to small changes in the analysis. This is exemplarily shown in \cref{fig:leakage}, which compares the intensity spectra of the \WaveK0-0+{K^{*}(892)}\pi P wave obtained from various systematic studies.

By performing detailed studies of our analysis formalism, we could demonstrate that this loss of distinguishing power affects only an identifiable and limited subset of partial waves and is mainly limited to the range $\mKpipi \lesssim \SI{1.5}{\GeVcc}$. These partial waves and mass ranges cannot be interpreted in terms of physics signals. This unfortunately limits, for example, our access to the $K^*(892) \pi$ decay of $\JP=0^-$ states, which would be important to study in the light of our searches for a potential exotic strange meson discussed in \cref{sec:exotic}.
However, those partial waves and mass ranges that are unaffected by the loss of distinguishing power can still be interpreted in terms of physics signals (see ref.~\cite{Wallner:2022prx} for details).

%=============================================================================
%=============================================================================
%=============================================================================
\section{High-Precision Strange-Meson Spectroscopy at AMBER}

To overcome the limitations for strange-meson spectroscopy at COMPASS discussed above is one of the main goals of the strange-meson spectroscopy program proposed by AMBER.
AMBER is a new QCD facility that will be located at CERN's M2 beam line, where the COMPASS experiment has been located so far. The proposal for phase 1 of AMBER mainly focuses on physics with high-energy muon, pion, and proton beams~\cite{Adams:2676885}. AMBER phase 1 has been approved by CERN RRB and the reference number NA66 is assigned to the experiment. These measurements are currently in preparation.

Phase 2 of AMBER focuses on physics with high-energy kaon beams~\cite{Adams:2018pwt}. A high-precision measurement of the strange-meson spectrum is one of the proposed measurements. The goal is to obtain a kaon diffraction data set that is about ten times larger than the one obtained by COMPASS.
One of the key requirements for this program is the upgrade of the final-state particle identification to cover a wider momentum range with a large acceptance. Other requirements are an efficient beam-particle identification, high-resolution tracking, and efficient photon detection to access also final states with neutral particles. These improvements of the experimental acceptance would eliminate the analysis limitations discussed in \cref{sec:leakage} and, at the same time, increase the size of the data set.

The size of the data set for strange-meson spectroscopy could be increased further by increasing the fraction of kaons in the beam using the radio-frequency separation technique. This technique is based on the time of propagation of the beam particles between two radio-frequency cavities, which is different for kaons and pions (see ref.~\cite{Adams:2018pwt} for details). The applicability of this technique to the M2 beam line is currently under study.

Obtaining such a large data set for strange-meson spectroscopy would not only lead to unprecedented precision in the measured properties of known states, but would also allow us to search for potential new signal in the data that are too weak to be observed by previous experiments.
In addition, such a large data set would allow us applying novel analysis methods. A prime example for such an analysis method is the so-called freed-isobar partial-wave analysis~\cite{Adolph2015,Krinner2018a}. This method extracts the amplitudes of selected two-body subsystems in a three-body final state, such as the $K^- \pi^+$ subsystem in the \Kpipi final state. Using this method, we can extract additional information from the data and, at the same time, reduce the model bias compared to a conventional partial-wave analysis. However, this method introduces much more free parameters in the partial-wave model and hence requires large samples to constrain these free model parameters. 

In summary, the strange-meson spectrum still holds many open questions and discovery opportunities. Many candidates for strange mesons require further confirmation and the search for exotic strange mesons has just started. At COMPASS, we collected the so far world's largest sample on diffractively produced \Kpipi events, based on which we performed the currently most detailed and comprehensive partial-wave analysis of this final state. However, the COMPASS analysis is limited by the limited coverage of the final-state particle identification and by the small fraction of kaons in the beam.
The goal of strange-meson spectroscopy measurements at AMBER is to overcome these limitations and to rewrite the PDG listings for strange mesons based on a single self-consistent analysis that covers the full mass range.

\bibliography{references}

\end{document}